\documentclass[aps,12pt,prb]{revtex4}
\usepackage{enumerate}
\usepackage{amsmath}

\begin{document}
\baselineskip18pt

\title{Information dynamics: Temporal behavior of  uncertainty measures}
\author{Piotr Garbaczewski\thanks{electronic address pgar@uni.opole.pl}}
\affiliation{Institute of Physics, University of Opole, 45-052
Opole, Poland}
\begin{abstract}
We carry out a systematic study of uncertainty measures that are
generic to  dynamical processes of varied origins, provided they
induce suitable continuous probability distributions. The major
technical  tool are the information theory methods and inequalities
satisfied by Fisher and Shannon information measures. We focus on a
compatibility of these inequalities with the prescribed
(deterministic, random or quantum) temporal behavior of
 pertinent probability densities.
\end{abstract}
\maketitle \noindent {\bf Keywords:} Information functionals;
information theoretic inequalities;  Shannon entropy;  Fisher
information; dynamics of probability densities; entropy methods;
Smoluchowski processes;
Schr\"{o}dinger picture evolution.
\vskip1.0cm
\section{Information and uncertainty}

Our primary motivation is  an information-theoretic conceptual
background and methods of  analysis adopted  for  probability
distributions.
 Such notions like  information,  uncertainty, indeterminacy and/or
  information deficit are naturally quantified in terms of information inequalities.
  A relationship between statistical (informational) and thermodynamic
notions of entropy of a physical (model) system, has
 received some  attention  as well,\cite{shannon,cover,ohya,stam,dembo}.

The pertinent  information  measures are seldom considered in the
time domain. Our  main purpose is to quantify  their temporal
behavior, while taking for granted that dynamical processes of
interest do induce suitable continuous probability densities, see
e.g.\cite{sobczyk,gar,gar1,mackey}. A particular attention is paid
to the time evolution of information entropies  and inferred
 uncertainty measures\cite{sobczyk,gar,mackey}.

\subsection{Entropic functionals}

To facilitate further  discussion, we  shall not attempt a fully
fledged space-time formalism, and pass to  time-dependent model
systems in one space dimension.  Let us consider continuous
probability densities on the real line, with or without an explicit
time-dependence: $\rho \in L^1(R); \int_R \rho (x) \, dx =1$.  Our
minimal demand is that the first and second moments of
 each density are finite.

 Therefore  we   can introduce a two-parameter family $\rho _{\alpha ,\sigma }(x)$,
 labeled by the mean value $ \langle x\rangle = \int x\, \rho (x)\,  dx=\alpha \in R $ and  the standard deviation
 (here, square root of the variance) $\sigma \in R^+ $, $\sigma ^2 = \langle (x- \langle x\rangle )^2\rangle $.

 For a given probability density $\rho $ we name a
function $- \ln \rho (x)$ a \it surprise  level \rm function  and
identify its mean value    with the familiar notion  ${\cal{S}}(\rho )$ of the Shannon
entropy of a continuous probability distribution, c.f.\cite{gar}:
\begin{equation}
{\cal{S}}(\rho )=  - \langle \ln \rho \rangle = - \int \rho (x) \ln
\rho (x)  dx\, .
\end{equation}
Let us assume $\rho (x) $ to be (weakly) differentiable, so that we
can give meaning to its first, second and third derivatives.

Besides an obvious information-theoretic notion of the Shannon
entropy, we introduce another information-theory functional
${\cal{F}} (\rho )$, often named  the Fisher information measure
(the name originates from the statistical inference theory):
\begin{equation}
{\cal{F}} (\rho ) \doteq  \langle ({\nabla \ln \rho })^2\rangle =
\int {\frac{(\nabla \rho )^2}{\rho }}  dx \, .
\end{equation}

The above introduced expressions   $\ln \rho (x)$ and $\nabla \ln
\rho $ allow to infer a number of interesting formulas.  We assume
the natural boundary data at finite  or  infinite (employed in
below) integration boundaries. We have:
\begin{equation}
-  \nabla \ln \rho =  - {\frac{\nabla \rho }\rho }  \Longrightarrow
- \langle {\frac{\nabla \rho }\rho }\rangle =0
\end{equation}
 and next
\begin{equation}
- \Delta  \ln \rho = -
{\frac{\Delta \rho }\rho } + {\frac{(\nabla \rho )^2}{\rho ^2}}  \Longrightarrow
 -\langle \Delta \ln \rho \rangle = \langle
{\frac{(\nabla \rho )^2}{\rho ^2}}\rangle = \langle (\nabla \ln \rho )^2\rangle  \, .   \label{por}
\end{equation}

The following identities  hold true:
\begin{equation}
- {\frac{\Delta \rho ^{1/2}}{\rho ^{1/2}}}  = {\frac{1}2}
[- {\frac{\Delta \rho }\rho }  + {\frac{1}2} {\frac{(\nabla \rho )^2}{\rho ^2}}] \Longrightarrow
 \nabla ({\frac{\Delta \rho ^{1/2}}{\rho ^{1/2}}}) =   {\frac{1}{2\rho }}   \nabla (\rho \Delta \ln \rho )  \label{fisher0}
 \end{equation}
where we encounter a  potential for a    "Newton-type  force field". Its  functional formula should  be compared with Eq.~(\ref{por}).
The    mean value  of the potential function is non-negative
 \begin{equation}
-  \langle  {\frac{\Delta \rho ^{1/2}}{\rho ^{1/2}}} \rangle = {\frac{1}4} \langle
{\frac{(\nabla \rho )^2}{\rho ^2}}\rangle  = - {\frac{1}4}   \langle \Delta \ln \rho \rangle  \, ,
\end{equation}
while this  of the related "force"  vanishes:
\begin{equation}
 \langle \nabla ({\frac{\Delta \rho ^{1/2}}{\rho ^{1/2}}}) \rangle = 0 \, .
\end{equation}

In the above systematics of derivatives   there was no indication of  a  specific  physical context. Nonetheless  a number of
physically interesting quantities can be immediately recognized.
They do notoriously appear in the local conservation laws for  diffusion-type processes  and in the
 hydrodynamical formulation of the  Schr\"{o}dinger picture  quantum dynamics\cite{gar,gar1}.

 Namely, while  keeping in mind (hitherto disregarded) dimensional coefficients, we realize   that    Eq.~(3) introduces a functional
 expression for an osmotic velocity  field.   Eqs.~(4) and (5) actually  set links
  between the hydrodynamical-type  pressure function\cite{gar1}
  and the Fisher functional.
  Eqs.~(6)  and (7) relate the so-called quantum potential and the pressure.   Eq.~(8) demonstrates   that the mean value of a
  (known as quantum) potential
 and the value of the  pressure functional do coincide, while Eq.~(8)  tells us  that the mean value of the
  inferred (quantum) force necessarily vanishes.

 Let us emphasize that our only input was  a  \it surprise level function \rm $- \ln \rho (x)$  for a continuous probability
density $\rho $ on $R$,
 admitting first and second moments, with suitable  differentiability properties  and  natural boundary data being  implicit.
 No specific   physical motivations  (like e.g.  quantum  or random dynamics)  were spelled out.

\subsection{Information inequalities}

Let us  consider a one-parameter $\alpha $-family of densities
whose  mean square deviation
 value   is   fixed at  $\sigma $. We have
\begin{equation}
{\cal{S}}(\rho )\leq (1/2) \ln (2\pi e\sigma ^2) \label{Fisherin}
\end{equation}
with a maximum
 for  a Gaussian  probability  density with  the  prescribed fixed   standard deviation $\sigma
 $.  By introducing the mean value:
\begin{equation}
\langle [\sigma ^2\nabla \ln \rho  + (x- \langle x\rangle )]^2\rangle \geq 0
\end{equation}
we readily arrive at an inequality
\begin{equation}
{\cal{F}} (\rho ) \geq {\frac{1}{\sigma ^2}}  \label{inequality}
\end{equation}
in which a minimum  of ${\cal{F}}$ is achieved   (among all
densities with a fixed value of $\sigma $)  if and only if $\rho $
is a $\sigma $-Gaussian, that in parallel with a  maximum for
${\cal{S}}$,  compare e.g.\cite{stam,furth}.

We stress that the above information  inequalities
Eq.~(\ref{Fisherin}) and (\ref{inequality}) set respectively lower
and upper bounds upon Fisher and Shannon functionals, while
evaluated with respect to  any  density  in the set of all
admissible ones (e.g.  with  once fixed  for all standard deviation
value $\sigma $).

Although we have carefully avoided any impact of dimensional
quantities, the above Eq.~(\ref{inequality}) actually associates a
primordial
"momentum-position"  indeterminacy relationship (here, devoid of any quantum connotations)  with  the
probability distributions  under consideration.  Namely, let $D$
be a positive diffusion constant with dimensions of $\hbar  /2m$
or $k_BT/m\beta $,  c.f.\cite{gar}.
 We define an osmotic velocity field
 $u=u(x)=  D\nabla \ln \rho $. There holds:
 \begin{equation}
 \Delta x \cdot \Delta u  \geq D     \label{brown}
 \end{equation}
which  correlates  the  position variance  $\Delta x=  \langle [x-\langle x\rangle ]^2\rangle ^{1/2}$ with the  osmotic velocity variance
$\Delta u = \langle [u - \langle u \rangle ]^2\rangle ^{1/2} $.

This property extends to time-dependent situations  and is known to be  respected by diffusion-type processes\cite{viola}.  Its primary version
for the free Brownian motion  has been found by R. F\"{u}rth\cite{furth}.

 As well, not
accidentally, the above  formula closely mimics, and in fact induces   (through  a reasoning based on probabilistic arguments)
  the  fully-fledged quantum mechanical position-momentum relationship
$\Delta x \cdot  \Delta p \geq \hbar /2$,  (tentatively replace $u$ by $mu$ and set $\hbar /2$ instead  of $D$, see e. g.\cite{golin}.
 We shall   come back to  this point in below.

To conclude this section, let us point out that, given  $\rho (x)$ and a  suitable   function $f(x) $, we can
 generalize the  previous arguments.  Let us introduce notions of a variance and covariance (here, directly borrowed from the random
 variable analysis\cite{golin}) for $x$ and $f(x)$. By means of the Schwarz inequality, we get:
 \begin{equation}
  \langle [x- \langle x\rangle ]^2\rangle \cdot \langle [f - \langle f\rangle ]^2\rangle \geq
  (\langle [x- \langle x\rangle ]\cdot \langle [f - \langle f\rangle ]\rangle )^2  \, ,
  \end{equation}
  hence, accordingly
  \begin{equation}
 Var(x) \cdot  Var(f) \geq     Cov ^2(x,f) \, .
 \end{equation}
 We note that for an osmotic velocity field $u(x)$, we have $\langle u\rangle =0$ and $\langle x\cdot u\rangle = - D$. Therefore
 \begin{equation}
  Var (x)\cdot Var(u) \geq  Cov^2(x,u) = D^2 \, ,
  \end{equation}
    as anticipated in Eq.~(\ref{brown}).

The casual intuition behind  physics-motivated  indeterminacy
relations is that of the Fourier transform. Indeed, for functions in
$L^2(R)$ a non-zero function and its Fourier transform  cannot be
both sharply localized.

Let us point out  remarkable information-theory
inequalities\cite{stam,ohya}:
  \begin{equation}
{\cal{F}}(\rho ) \geq (2\pi e)  \exp [-2{\cal{S}}(\rho )] \geq
1/\sigma ^2\, .  \label{chain}
\end{equation}
and note that an explicit  Fourier transformation input allows to
set an upper bound in this  chain of inequalities.

The crucial  step is to  disentangle the $L^2(R)$  ingredients in
the $L^1(R)$  functional form of the density $\rho $. This can be
accomplished on many ways. the simplest is either a multiplicative
decomposition $\rho (x)= \rho ^{1/2}(x) \cdot \rho ^{1/2}(x)$, where
clearly $\rho ^{1/2} \in L^2(R)$. More general  choice involves a
complex function $\phi \in L^2(R)$ and its complex conjugate $\phi
^* $ so that $\rho = \phi \cdot  \phi ^* =  |\phi |^2$.  (We recall
that the Fourier transform of a real function typically is a complex
function.)

Given  an $L^2(R)$-normalized function  $\psi (x)$. We denote $({\cal{F}}\psi )(p)$ its Fourier
transform. The corresponding probability densities follow:
 $\rho (x) = |\psi (x)|^2$ and $\tilde{\rho }(p) = |({\cal{F}}\psi )(p)|^2$.\\
We introduce the related position and momentum  information  (differential, e.g. Shannon) entropies:
\begin{equation}
{\cal{S}}(\rho )\doteq S_q =  - \langle \ln \rho \rangle = - \int \rho (x) \ln \rho (x)  dx
\end{equation}
    and
    \begin{equation}
    {\cal{S}}(\tilde{\rho })\doteq
S_p= - \langle  \ln \tilde{\rho }\rangle =   - \int \tilde{\rho }(p) \ln \tilde{\rho }(p)  dp
\end{equation}
 where  ${\cal{S}}$ denotes the  Shannon
 entropy for a continuous probability  distribution.  For the sake of clarity, we
 use dimensionless quantities, although there exists a consistent procedure for handling
  dimensional quantities in the Shannon entropy definition.

We assume both entropies to take finite values. Then, there holds
the familiar   entropic uncertainty relation which is the sole
consequence of the Fourier transform properties in $L^2(R)$
\cite{mycielski}:
\begin{equation}
S_q + S_p \geq  (1 + \ln \pi ) \, .  \label{uncertainty}
\end{equation}

 Let us notice that in view of properties of  the Fourier transform, there is a complete symmetry between  the
 inferred  information-theory
 functionals. After the Fourier transformation, the Parceval identity implies that  the chain of
  inequalities  Eq.~(\ref{chain}) can be  faithfully reproduced (while replacing $\rho $  by $\tilde{\rho }$) for
 the "momentum -space" density $\tilde{\rho }$ with the  variance $\tilde{\sigma }^2$.  As a consequence, taking into   account
 the entropic uncertainty relation Eq.~(\ref{uncertainty}), we arrive at\cite{mycielski}:
\begin{equation}
4 \tilde{\sigma }^2 \geq  2(e\pi )^{-1}  \exp[-2  \langle \ln \tilde{\rho }\rangle ] \geq  (2e\pi ) \exp[ 2 \langle \ln  \rho \rangle ]
\geq \sigma ^{-2} \label{chain1}
\end{equation}
which adds a Fourier transform-inferred  upper bound to the previous
inequalities Eq.~ (\ref{chain})
\begin{equation}
 4 \tilde{\sigma }^2 \geq   {\cal{F}}(\rho ) \geq
1/\sigma ^2\,
\end{equation}
and sets related  upper and lower bounds upon the Shannon entropy as
well.

All that has been considered  with no mention of any time evolution.
Since the  dynamics has a physical provenance,  we should carefully
investigate the $\rho $-factorization issue and an impact of a
priori given dynamical rules for $\rho _0(x) \rightarrow \rho (x,t)$
 upon its concrete realisation.  We shall focus on the standard
random dynamics (Smoluchowski processes and  phase-space motion) and
the Schr\"{o}dinger picture quantum dynamics.

\section{Quantum indeterminacy in the time domain}

If  following  conventions we define the squared  standard deviation
value  for an observable $A$ in a pure state $\psi $ as $(\Delta
A)^2 = (\psi , [A - \langle A\rangle ]^2 \psi )$ with $\langle A
\rangle  = (\psi , A\psi)$, then for the  position $X$ and momentum
$P$ operators we have the following version of the entropic
uncertainty relation  (here expressed through so-called entropy
powers, see e.g.   \cite{ohya}, $\hbar \equiv 1$):
\begin{equation}
\Delta X \cdot \Delta P \geq  {\frac{1}{2\pi e}} \,
 \exp[{\cal{S}}(\rho )  + {\cal{S}}(\tilde{\rho })] \geq {\frac{1}{2}} \,  \label{un}
\end{equation}
which is  an alternative version of the entropic uncertainty
relation.
  For Gaussian densities,   $ (2\pi e )\Delta X \cdot \Delta P =
 \exp[{\cal{S}}(\rho )  + {\cal{S}}(\tilde{\rho })]$ holds true, but the     minimum $1/2$
  on the right-hand-side of Eq.~(\ref{un}),   is  not necessarily  reached.

Let us consider a momentum operator $P$ that is conjugate  to the
position operator  $X$ in the adopted dimensional convention $\hbar
\equiv 1$. Setting $P= - i d/dx$ and presuming that all  averages
are finite,  we get:
\begin{equation}
[\langle P^2\rangle - \langle P\rangle ^2] =  (\Delta P)^2=
\tilde{\sigma }^2 \, .
\end{equation}
The  standard indeterminacy relationship
 $\sigma  \cdot \tilde{\sigma }\geq (1/2)$  follows.

  In the above, no explicit time-dependence
has been  indicated, but all derivations go through with
any wave-packet  solution $\psi (x,t)$ of the Schr\"{o}dinger equation. The  induced dynamics of
 probability densities may imply the  time-evolution of  entropies: $S_q(t), S_p(t)$ and thence the dynamics of
 quantum  uncertainty measures  $ \Delta X  (t) =  \sigma (t)$ and $ \Delta P(t)= \tilde{\sigma }(t) $.

We consider the Schr\"{o}dinger equation in  the form:
\begin{equation}
i\partial _t  \psi  = - D  \Delta  \psi   +
{\frac{{\cal{V}}}{2mD}} \label{Schroedinger} \psi \, .
\end{equation}
where the  potential ${\cal{V}}= {\cal{V}}(\overrightarrow{x},t)$ (possibly time-dependent)  is  a  continuous
(it is useful, if bounded from below)  function  with dimensions of
 energy, $D=\hbar /2m$.

By employing the  Madelung decomposition:
\begin{equation}
 \psi = \rho ^{1/2} \exp(is/2D) \, ,
 \end{equation}
   with the
phase function $s=s(x,t)$  defining  a (current)  velocity field  $v=\nabla s $,  we readily  arrive at the continuity equation
\begin{equation}
\partial _t \rho = - \nabla (v \rho )
\end{equation}
 and the generalized Hamilton-Jacobi equation:
\begin{equation}
\partial _ts +\frac{1}2 ({\nabla }s)^2 + (\Omega -  Q) = 0 \label{jacobi1}
\end{equation}
where  $\Omega = {\cal{V}}/m$ and,   after  introducing an
 osmotic velocity field $
u(x,t) = D\nabla \ln \rho (x,t)$ we have, compare e.g. our
discussion of Section I:
\begin{equation}
Q  =  2D^{2}{\frac{\Delta \rho ^{1/2}} {\rho ^{1/2}}} = {\frac{1}2}
u^2 + D \nabla \cdot u \, .
\end{equation}

If  a quantum mechanical expectation value of the standard Schr\"{o}dinger  Hamiltonian  $\hat{H}= -(\hbar ^2/2m) \Delta + V$
  exists (i.e.  is finite\cite{gar2}),
  \begin{equation}
 \langle \psi | \hat{H}|\psi \rangle \doteq E < \infty
 \end{equation}
  then the unitary  quantum dynamics warrants that this value is a  constant
 of the Schr\"{o}dinger picture evolution:
\begin{equation}
{\cal{H}} = {\frac{1}2} [\left< {v}^2\right> + \left< {u}^2\right>]  +
\left<\Omega \right>  =  - \left< \partial _t s\right>  \doteq {\cal{E}} = {\frac{E}m}= const  \, . \label{total}
\end{equation}
Let  us notice that  $\langle u^2\rangle = - D \langle \nabla u \rangle $ and therefore:
\begin{equation}
{\frac{D^2}{2}} {\cal{F}} =   {\frac{D^2}{2}} \int {\frac{1}{\rho }}
\left({\frac{\partial \rho }{\partial x }} \right)^2\, dx  =   \int
\rho \cdot  {\frac{u^2}{2}} dx = -   \langle Q \rangle \, .
\label{Fisher1}
  \end{equation}

We  observe that  $D^2{\cal{F}}$  stands for  the  mean square deviation  value  of a function $u(x,t)$ about its mean value
$\langle u \rangle =0$, whose vanishing is a consequence of the boundary conditions (here, at infinity):
\begin{equation}
(\Delta u)^2 \doteq \sigma _u^2 =  \langle [u- \langle u\rangle ]^2\rangle = \langle u^2\rangle  = D^2 {\cal{F}} \, .
\end{equation}
The  mean square deviation of $v(x,t)$ about its mean value $\langle v\rangle$ reads:
\begin{equation}
(\Delta v)^2 \doteq \sigma ^2_v = \langle v^2 \rangle -
\langle v \rangle ^2\, .
\end{equation}
It is clear, that with the definition $P= -i(2mD) d/dx$, the mean value of the operator $P$ is related to the mean value of a
function $v(x,t)$  (we do not discriminate between  technically  different implementations  of the mean):  $\langle P\rangle = m\langle v \rangle $.
Accordingly,
\begin{equation}
\tilde{\sigma }^2  = (\Delta P)^2 = \langle P^2\rangle - \langle P\rangle ^2
\end{equation}

Moreover, we can directly check that with $\rho = |\psi |^2$ there
holds\cite{hall}:
\begin{equation}
{\cal{F}}(\rho ) = {\frac{1}{D^2}} \sigma ^2_u = \int dx |\psi |^2[\psi '(x)/\psi (x) + {\psi ^*}'(x) /\psi ^*(x)]^2=   \label{ha}
\end{equation}
$$
4\int dx {\psi '}^*(x) {\psi '}(x)  + \int dx |\psi (x)|^2 [ \psi '(x)/\psi (x) -{\psi ^*}'(x) /\psi ^*(x)]^2  =
$$
$$
{\frac{1}{m^2D^2}} [ \langle P^2\rangle - m^2 \langle v^2 \rangle ] = {\frac{1}{m^2D^2}}[ (\Delta P)^2 -  m^2 \sigma ^2_v]
$$
i.e.
\begin{equation}
 m^2(\sigma ^2_u + \sigma ^2_v) = \tilde{\sigma }^2 \, . \label{heisenberg}
\end{equation}
  It is interesting to notice that $\langle (P - mv )\rangle  =0$
and the corresponding mean square deviation  reads: $
 \langle (P-mv)^2\rangle = \langle P^2\rangle - m^2\langle v^2\rangle = m^2D^2
 {\cal{F}}$.

By passing to dimensionless quantities in Eqs.~(\ref{ha}) (e.g.
$2mD\equiv 1$), and denoting $p_{cl} \doteq
 (\arg \,  \psi (x,t) )' $ we get:
\begin{equation}
{\cal{F}} = 4[\langle P^2\rangle - \langle p^2_{cl}\rangle ] = 4[(\Delta P)^2 -  (\Delta p_{cl})^2] = 4[\tilde{\sigma }^2 - \tilde{\sigma }_{cl}^2]
\end{equation}
and therefore  the chain of inequalities Eq.~(\ref{chain}) gets a sharper form:
 \begin{equation}
4\tilde {\sigma }^2 \geq 4[\tilde{\sigma }^2 - \tilde{\sigma }_{cl}^2] = {\cal{F}}      \geq (2\pi e)  \exp [-2{\cal{S}}(\rho )]
\geq  {\frac{1}{\sigma ^2}} \, .
\end{equation}
We recall that all "tilde" quantities can be  deduced from the  once  given $\psi $ and  its Fourier transform $\tilde{\psi}$.

   As  a side comment let us add that a direct consequence of the  mean  energy conservation law Eq.~(\ref{total}) are identities:
   $\langle P^2\rangle /2m = E - \langle {\cal{V}}\rangle $
   and
    \begin{equation}
    {\cal{F}} =  {\frac{1}{m^2D^2}} [ \langle P^2\rangle - m^2 \langle v^2 \rangle ] =
    {\frac{1}{D^2}} [2({\cal{E}} - \langle \Omega \rangle ) - \langle v^2 \rangle ]
\end{equation}
plus a complementary expression for the variance of the momentum
observable:
\begin{equation}
  (\Delta P)^2= 2 m  (E - \langle [{\frac{m}2}  \langle v \rangle ^2  + {\cal{V}}]\rangle
  )\, .
  \end{equation}
That combines into the chain of   inequalities between various energy characteristics:
\begin{equation}
 E -\langle
{\cal{V}}\rangle
> m\langle v^2\rangle /2 \geq  m \langle v\rangle ^2/2\geq 0\, .
\end{equation}

\section{Indeterminacy relations for  diffusion-type processes}

Let us consider  spatial random motions, like  e.g. standard
 Smoluchowski processes and their generalizations.
Let us consider $\dot{x} = b(x,t) +A(t)$ with $\langle A(s)\rangle =0 \, , \,  \langle A(s)A(s')\rangle
= \sqrt{2D} \delta (s-s')$   and the corresponding Fokker-Planck equation for the probability density $\rho $
which we analyze under the natural boundary conditions:
\begin{equation}
\partial _t\rho =
D\triangle \rho -  \nabla \cdot ( b \rho ) \label{fokker}
\end{equation}
which we analyze under the natural boundary conditions.

We assume the gradient form for the forward drift  $b= b(x,t)$ and take $D$ as a  diffusion   constant with dimensions of
$k_BT/m\beta $.
 By introducing $u(x,t) = D \nabla \ln \rho (x,t)$ we define the current velocity of the process
 $v(x,t) = b(x,t) - u(x,t)$, in terms of which the  continuity equation  $\partial _t \rho = - \nabla (v\rho )$ follows.
The diffusion current reads $j=v \rho $.

As mentioned before, we have an obvious  indeterminacy relationship for the osmotic velocity field
   $Var (x)\cdot Var(u) \geq  Cov^2(x,u) = D^2$. The corresponding relationship for the current velocity field
$Var(x) \cdot  Var(f) \geq     Cov ^2(x,f)$,  contrary to the previous quantum   reasoning,  does not naturally yield
 any analogue   of the  Heisenberg-type position-momentum uncertainty formulas, c.f. Eq.~(35).

 The  cumulative identity
 $Var (x)\cdot [Var(u) + Var(v)] \geq Cov^2(x,v) + D^2$, reproduced in Ref.\cite{golin}, does not convey any useful message
 about  the diffusion process. It  cannot be  directly  inferred from the Fisher functional   ${\cal{F}}(\rho )$
 which  actually \it    was \rm  the case  in our previous, quantum  discussion, e.g. where we have had $Var P=
 Var(mu) + Var(mv)$. For spatial diffusion processes, the latter identity  is plainly  nonexistent, since there
 is no diffusive analogue of  the quantum momentum  observable.

Let us mention  an early   attempt\cite{shin}  to set an uncertainty principle for general diffusion processes.
 If adopted to our convention
(natural boundary data), in view of $\langle u\rangle =0$ and $v= b- u$, we have $\langle v\rangle = \langle b \rangle $.

For an arbitrary real constant $C\neq 0$, we obviously have:  $[C \cdot ( v- \langle v\rangle ) + (x-\langle x\rangle ]^2 \geq 0$.
The mean value of this auxiliary inequality  reads:
\begin{equation}
C^2 (\Delta v) ^2  + 2C[\cdot  Cov(x,b) +   D] +  (\Delta x)^2 \geq 0\, .
\end{equation}
and is  non-negative for all $C$,  which  enforces a condition
\begin{equation}
[D + Cov(x,b)]^2 - (\Delta v) ^2\cdot (\Delta x)^2   \leq 0 \, .
\end{equation}
Note that  $Cov(x,v)= D + Cov(x,b)$, so we have in fact  an alternative derivation  of the previous indeterminacy
relationship $Var(x) \cdot Var(v) \geq Cov^2(x,v)$
    for the current velocity field.

 In case of  Smoluchowski processes, forward drifts are
      proportional to externally imposed force fields, typically through $b =F/m\beta $. Therefore  the position-current
       velocity      dispersion correlation  is controlled by  $Cov(x,F)$. For  the free  Brownian motion (e.g. the Wiener process)
     we have $b=0$, and hence $Cov(x,v)=D$.

To get a deeper insight into the "position-momentum indeterminacy issue" for diffusion processes, let us begin
from a classic observation that, once we set  $b= - 2D\nabla \Phi $  with $\Phi = \Phi (x)$,  a substitution:
\begin{equation}
\rho (x,t) \doteq \theta _*(x,t) \exp [- \Phi (x)]
\end{equation}
with $\theta _*$ and $\Phi $ being real functions, converts the Fokker-Planck equation Eq.~(\ref{fokker})
 into a generalized diffusion  equation for $\theta _*$:
 \begin{equation}
 \partial _t \theta _* = D \Delta \theta _* -  {\frac{{\cal{V}}(x)}{2mD}}\theta _*
 \end{equation}
 and its time adjoint
\begin{equation}
\partial _t \theta = -D\Delta \theta  + {\frac{{\cal{V}}(x)}{2mD}}\theta
\end{equation}
  for a real function $\theta (x,t) = \exp [- \Phi (x)]$,
where
\begin{equation}
{\frac{{\cal{V}}(x)}{2mD}} =  {\frac{1}2} ({\frac{b^2}{2D}} + \nabla \cdot b) =  D[(\nabla \Phi )^2 -\Delta \Phi ] \, . \label{fokkerpot}
\end{equation}
Let us note an obvious factorization property for the Fokker-Planck probability density:
\begin{equation}
\rho (x,t) = \theta (x,t)  \cdot \theta _*(x,t)
\end{equation}
which stays in affinity with  a quantum mechanical factorization formula $\rho =  \psi ^*  \psi $,  albeit presently
realized in terms of two  real functions $\theta $ and $\theta ^*$, instead of a complex conjugate pair.

Let us mimic basic steps, outlined in Eq. (34) for the complex factorization of $\rho $,  but in terms of two real functions
$\theta $ and $\theta _*$.  We have:
\begin{equation}
{\cal{F}}(\rho ) = {\frac{1}{D^2}} \sigma ^2_u = \int dx (\theta \theta ^*) [{\frac{\theta '}{\theta }}  +
{\frac{{\theta _*}'}{\theta _*}}]^2= \label{ha1}
\end{equation}
$$
4\int dx {\theta '}_* {\theta '}  + \int dx (\theta \theta _*)[{\frac{\theta '}{\theta }}  -
{\frac{{\theta _*}'}{\theta _*}}]^2 \, .
$$

Since a continuity equation  $\partial _t \rho = - \nabla j$ is identically fulfilled by
\begin{equation}
j(x,t) = \rho (x,t) v(x,t) = D(\theta _* \nabla \theta - \theta \nabla \theta _*)
\end{equation}
we obviously get:
\begin{equation}
{\cal{F}}(\rho )=  {\cal{F}}(\rho =\theta \theta _*)= 4 \int dx (\nabla \theta )(\nabla \theta ^*)
  + {\frac{1}{D^2}} \langle v^2\rangle  =  -{\frac{2}{mD^2}} \langle  {\cal{V}}\rangle + {\frac{1}{D^2}} \langle v^2\rangle  \label{fisher}\, ,
\end{equation}
 to be compared with the quantum mechanical result:
\begin{equation}
 {\cal{F}}(\rho =|\psi |^2)= 4 \int dx (\nabla \psi )(\nabla \psi ^*)  -{\frac{1}{D^2}} \langle v^2\rangle  =
    {\frac{1}{D^2}} [2({\cal{E}} - \langle \Omega \rangle ) - \langle v^2 \rangle ]
\, .   \label{qfisher}
\end{equation}
By reintroducing  $\Omega = {\cal{V}}/m$ in Eq.~(\ref{fisher}):
\begin{equation}
{\cal{F}}(\rho =\theta \theta _*)  ={\frac{1}{D^2}} [ - 2\langle \Omega \rangle  + \langle v^2 \rangle ] \label{fisher1}
\end{equation}
we achieve a notational  conformity with Eq.~(\ref{qfisher}).

The major difference between the formulas Eq.~(\ref{fisher1}) and
Eq.~(\ref{qfisher}), apart from the presence or absence of an
additive term ${\cal{E}} \in R$, is that a  diffusive potential
${\cal{V}}$ has a pre-determined functional form,
Eq.~(\ref{fokkerpot}).   Our general restriction on   ${\cal{V}}$,
irrespective of whether this potential enters the Schr\"{o}dinger or
the generalized heat equations,  is    that  it  should be a
continuous and bounded from below function\cite{zambrini}.   In the
diffusive case this demand guarantees that $\exp(-tH)$ with $H\doteq
-D\Delta + (1/2mD){\cal{V}}$  is a legitimate dynamical semigroup
operator.

Let us add that
\begin{equation}
{\cal{F}}(\rho =\theta \theta _*)= {\frac{2}{mD^2}}  \langle {\frac{mv^2}2} - {\cal{V}}\rangle \Rightarrow
\langle {\frac{mv^2}2} - {\frac{mu^2}2}- {\cal{V}}\rangle  =0  \label{mean}
\end{equation}
while
\begin{equation}
{\cal{F}}(\rho =|\psi |^2)={\frac{2}{mD^2}} [E  - \langle {\frac{mv^2}2}  + {\cal{V}} \rangle ] \Rightarrow
\langle {\frac{mv^2}2} + {\frac{mu^2}2}  + {\cal{V}} \rangle  = E \, . \label{mean1}
\end{equation}
The variances of osmotic and current velocity fields are correlated, respectively, as follows
\begin{equation}
\rho =\theta \theta _* \Longrightarrow  m^2[ (\Delta u)^2   - (\Delta v)^2] = 2m [ {\frac{m\langle v\rangle ^2}2} -
\langle {\cal{V}}\rangle ]         \label{diff}
\end{equation}
and
\begin{equation}
\rho = \psi \psi ^* \Longrightarrow   m^2[(\Delta u)^2  +(\Delta v)^2] = 2m [E  - ({\frac{m\langle v\rangle ^2}2}
 + \langle {\cal{V}} \rangle ] = (\Delta P)^2
\, .                                                           \label{quant}
  \end{equation}
Since  $(\Delta u)^2 \geq D^2/\sigma ^2$, in view of Eqs.~(57) and (\ref{heisenberg}) we readily  arrive at  the  standard
quantum indeterminacy relation for  position and momentum  observables $\Delta P\cdot \Delta X \geq mD$.

In case  of diffusion-type processes we definitely  encounter a
non-standard situation. On the left-hand-side of Eq.~(\ref{diff}),
there appears a difference of variances for   the  current  and
osmotic velocity fields, instead of their sum, like e.g. in
Eq.~(\ref{quant}).   This expression is not necessarily  positive
definite, unless $\langle {\cal{V}}\rangle \leq 0$ for all times.

Let us make a guess that  $\Delta u > \Delta v$, in the least
locally in time (in a  finite time interval). Then, the resulting
expression
\begin{equation}
 m^2(\Delta u)^2= m^2 \langle u^2\rangle =
  2m \langle {\frac{m v^2}2} - {\cal{V}}\rangle \doteq  (\Delta p_{u})^2 \geq {\frac{m^2D^2}{\sigma ^2}} \, ,
  \end{equation}
as we already know,  yields a dimensionally acceptable
position-momentum  indeterminacy relationship for  diffusion-type
processes,
\begin{equation}
  \Delta x \cdot  \Delta p_{u} \geq mD\, , \label{one}
\end{equation}
 where   $ \Delta p_{u} >0 $ may  be interpreted as the   pertinent    "momentum dispersion" measure.
  For the free Brownian motion we have ${\cal{V}} =0$  and $v= -u$, hence   Eq.~(\ref{brown}) is recovered.

Upon making  an opposite guess i. e.  admit $\Delta v >\Delta u$
(again. at least locally in  time),  in view of ${\cal{F}} \geq
1/\sigma ^2$, we  would have
\begin{equation}
 m^2 (\Delta v)^2  =
  m^2 (\Delta u)^2 +  2m [  \langle
{\cal{V}}\rangle   - {\frac{m\langle v\rangle ^2}2}]  \doteq (\Delta
p_v)^2 \geq {\frac{m^2D^2}{\sigma ^2}}
\end{equation}
and thus
\begin{equation}
 \Delta x \cdot \Delta p_v\geq mD  \, \label{two}
 \end{equation}
 would ultimately arise.

The above  two  indeterminacy options (\ref{one})  and  (\ref{two})
are a consequence of
 a possibly indefinite  sign  for a difference
$\Delta u - \Delta v$ of standard deviations, in the course of a
diffusion process. This sign issue seems  to   be a local in time
property and may  not persist in the asymptotic (large time) regime.
We shall give an argument towards a non-existence of a fixed
positive lower bound for the joint
 position-current velocity uncertainty
measure in the vicinity of an asymptotic stationary solution of  the
involved   Fokker-Planck equation.

In case of Smoluchowski diffusion processes we may take for granted
that they asymptotically approach\cite{mackey,gar} unique stationary
solutions, for which the current velocity $v$ identically vanishes.
Then $\Delta v =0$ as well, while   $0< Var(x) < \infty $ ( e. g.
$\Delta x $ stays finite).

In view of Eq. (\ref{mean}), an asymptotic value of  the  strictly
positive Fisher functional
 ${\cal{F}}$ equals $-(2/mD^2) \langle {\cal{V}}\rangle
>0$.  Accordingly, to secure ${\cal{F}}>0$,  an expectation value of ${\cal{V}}$ with respect
to the stationary probability density must be negative. Even, under
an assumption that
 ${\cal{V}}$  is bounded from below.

Consequently,  in the large time asymptotic we surely have $(\Delta
u)^2 >  (1/\sigma ^2)>   (\Delta v)^2$ and  $\Delta v \rightarrow
0$, while  $\sigma $ has a finite limiting value (an exception  is
the free Brownian motion when $\sigma $ diverges).
 The validity of the above argument can be checked
by inspection, after invoking an explicit solution for the
Ornstein-Uhlenbeck process\cite{gar,mackey}.

 Thus, $\Delta x \cdot \Delta p_v\geq mD$ does not hold true in the
vicinity of the asymptotic solution. On the contrary, $\Delta x
\cdot \Delta p_u\geq mD$ is universally valid.

\section{Entropy methods: Thermodynamical patterns of behaviour in diffusion-type processes}

\subsection{Thermodynamical hierarchy}

Diffusion processes  stand for an approximate description of  (macro)molecules whose motion is induced by a thermal environment.
As such they quantify  the  dynamics of  non-equilibrium thermodynamical systems.

The following  hierarchy of  thermodynamical systems is adopted  in below: {\it isolated} with no energy
and matter exchange with the environment,  {\it closed} with the energy   but  {\it no } matter exchange and
 {\it open} where energy-matter exchange  is unrestricted.
We keep in mind a standard text-book wisdom that all isolated systems evolve to the state of equilibrium
 in which  the entropy reaches its maximal value.  An approach towards equilibrium is  here interpreted as
 an approach towards most disorderly   state.

Our  further  attention  will be   focused on  non-isolated, albeit  {\it  closed},  random  systems and their somewhat different
  asymptotic features.
  Assuming  the  natural boundary data\cite{mackey,hasegawa,qian}, we shall introduce basic thermodynamical concepts and
 recall   the Helmholtz  extremum principle for an  intrinsically  random motion. Thermodynamic function(al)s, like e.g.
 an internal energy,  Helmholtz free energy and entropy  will be inferred,  through suitable averaging,  from
a priori prescribed   time-dependent  continuous probability densities.

  A concise resume of a non-equilibrium thermodynamics of {\it
  closed} systems comprises the  $I^{st}$  law  $\dot{U} = \dot{Q} +
  \dot{W}$  and  the $II^{nd}$  law $\dot{S}= \dot{S}_{int} +
  \dot{S}_{ext}$, where $\dot{S}_{int}\geq 0$ and $\dot{S}_{ext}  =
  \dot{Q}/T$, c.f. \cite{glansdorf,kondepudi}.
  We are fully aware that not all objects involved
  (like e.g. $Q$ can  viewed as legitimate  analogs of thermodynamic functions.
  Nonetheless, in the forthcoming discussion, the heat   exchange  and work time
  rates are always  well defined and  an issue
  of "imperfect differentials" is consistently bypassed.

  Thermodynamical extremum principles are usually invoked in
  connection with the large time  behavior of  irreversible
  processes.  Among a number of standard  principles,  for reference,
  we   recall  a specific one  named the Helmholtz
  extremum principle. If the temperature $T$ and  the available
  volume $V$ are kept constant, then
  the minimum of the Helmholtz free energy  $F= U - TS$  is
  preferred  in the course of the system evolution in time,
   and  there holds  $ \dot{F} = - T\dot{S}_{int} \leq
  0$.

    \subsection{Thermodynamics of  random phase-space motions}

    Let us  consider a  phase-space diffusion process governed
    by the Langevin equation
    $
    m\ddot{x}  + m\gamma \dot{x} = - \nabla V(x,t) + \xi(t)$,
    with standard assumptions about properties of the white noise:
    $\langle \xi (t)\rangle =0, \,   \langle \xi (t)\xi (t')\rangle =
    \sqrt{2m\gamma k_BT}\,  \delta(t-t')$.
    Accordingly, the pertinent phase-space density  $w=w(x,u,t)$ is a
    solution of the Fokker-Planck-Kramers equation with suitable
    initial data:
    \begin{equation}
    {\frac{\partial }{\partial t}}w(x,u,t) =
    \end{equation}
    $$
    \left[ - {\frac{\partial }{\partial x}} u + {\frac{\partial
    }{\partial u}} \left(\gamma u + {\frac{1}m} \nabla V(x,t)\right) +
    {\frac{\gamma k_BT}m} {\frac{\partial ^2}{\partial u^2}}\right] w
    $$
    Let us define  the  Shannon entropy $ {\cal{S}} = {\cal{S}}(t)$ of
    a continuous probability distribution  in the phase-space of the system:
    \begin{equation}
    {\cal{S}}(t) = - \int ( w \ln w) \, dx\, du = - \langle \ln w \rangle
    \end{equation}
    (By dimensional reasons we should insert a factor $h$ with physical
    dimensions of the action under the logarithm, i.e. use $\ln (h w)$
    instead of $\ln w$, but since we shall  ultimately  work with  time
    derivatives, this step  may be safely
    skipped.)

    An internal energy  $U$ of the  diffusion-type  stochastic process  we define as follows:   $U= \langle E\rangle  $, where
    $E=E(x,u,t) = {\frac{mu^2}2} + V(x,t)$. Then, the $I^{st}$ law   of thermodynamics   takes the form
    \begin{equation}
    \dot{\cal{Q}} + \dot{W} = \dot{U}
    \end{equation}
     where $\dot{W} \doteq  \langle
    \partial _t V\rangle$ is interpreted as the work externally performed upon the system. (For future reference we would like to
    stress a particular  importance of the time-dependent work term in quantum theory.)

    Furthermore, let us introduce an obvious analog  of the Helmholtz
    free energy:
    \begin{equation}F\doteq \langle E + k_BT \ln w\rangle =   U-TS
    \end{equation}
    so that
    \begin{equation}
    \dot{F} - \dot{W} = \dot{Q} - T\dot{S}     = - T\dot{S}_{int}
    \leq 0 \, .
    \end{equation}
    The above result  is a direct consequence of the Kramers equation.
    Under suitable assumptions concerning the proper behavior of
    $w(x,u,t)$ at $x,u$ integration boundaries (sufficiently rapid decay  at
    infinities) we have \cite{shizume}
    $
    \dot{\cal{Q}} = \gamma (k_BT - \langle mu^2 \rangle ) $ and
    $\dot{\cal{S}} = \gamma \left[ {\frac{k_BT}m} \langle
    \left({\frac{\partial \ln w}{\partial u}} \right)^2 \rangle  - 1\right]
    $.

    In view of $(1/T)\dot{Q} = \dot{S}_{ext}$,   the $II^{nd}$ law
    readily follows
    \begin{equation}
    \dot{\cal{Q}} - k_BT\dot{\cal{S}}=
    \end{equation}
    $$
     - {\frac{\gamma }m} \langle
    \left(k_BT{\frac{\partial \ln w}{\partial u}}
    + mu\right)^2\rangle  = -T\dot{S}_{int} \leq 0
    $$

     We denote $S \doteq k_B {\cal{S}}$ and so arrive at $\dot{Q}
    \leq  T\dot{S}$. As  a  byproduct of the discussion  we have  $\dot{F}
    \leq \dot{W}$.

     For time-independent $V=V(x)$, the
     extremum principle pertains to  minimizing the  Helmholtz free energy $F$  in the course of
      random motion:
    \begin{equation}
    \dot{F} = \dot{Q} - T\dot{S} \doteq -  T\dot{S}_{int}  \leq 0
    \end{equation}

    The preceding   discussion encompasses both the forced and   unforced (free) Brownian motion. When $V(x)\equiv 0$, then
     no asymptotic  state of  equilibrium (represented by a probability density) is accessible, the motion is sweeping.
    In the forced case    we  realize  that:
     $w_{*}(x,u) = {\frac{1}Z} \exp \left[ - {\frac{E(x,u)}{k_BT}}\right]$,
    is a stationary solution of the Krames equation. Therefore,  we   may expect that    the dynamics  actually  relaxes to  this a (unique)
     stationary state\cite{mackey} $w\rightarrow w_*$.
     Obviously, $w_{*} $  is
     non-existent in case of the free Brownian motion.

\subsection{Thermodynamics of  Smoluchowski processes}

Analogous  thermodynamical features are encountered in   spatial random motions, like  e.g. standard
 Smoluchowski processes and their generalizations. Given a  probability density $\rho(x,t)$ solution of a Fokker-Planck equation
$\partial _t\rho = D\triangle \rho -  \nabla \cdot ( b \rho )$.
The related  Shannon  entropy
${\cal{S}}(t)  = -\langle \ln \rho \rangle  $  typically is not a conserved quantity and, with boundary restrictions
that $\rho, v\rho, b\rho $ vanish  at spatial infinities or  finite integration interval borders,  various   equivalent
forms of the balance equation   follow.
We select, \cite{gar,qian}:
\begin{equation}
  D \dot{\cal{S}}  =  \left< {v}^2\right>
    -  \left\langle {b}\cdot {v}
 \right\rangle  \label{balance}  \, .
\end{equation}

A  thermodynamic formalism for Smoluchowski processes is straightforward.
We pass to time-independent drift fields
  and  set, while adjusting dimensional constants: $ b = \frac{f}{m\beta }$.
Exploiting  $j \doteq v\rho $,  $ f = - \nabla V $ and setting  $D=k_BT/m\beta $ we give
 Eq.~(\ref{balance}) a conspicuous  form of:
\begin{equation}
\dot{\cal{S}} = \dot{\cal{S}}_{int} + \dot{\cal{S}}_{ext}
\end{equation}
where $
k_BT \dot{\cal{S}}_{int}  \doteq m\beta   \left<{v}^2\right> \geq 0$
stands for the  entropy production, while
$ k_BT \dot{\cal{S}}_{ext} =   \dot{\cal{Q}} =  -  \int {f} \cdot {j}\,  dx =
- m\beta  \left\langle {b}\cdot {v}
 \right\rangle
$
 (as long as negative  which is not a must)  may be  interpreted as the   heat dissipation rate:
 in view of  $\dot{{\cal{Q}}}= - \int {f}\cdot {j}\,  dx$,
 there is  a definite  power release involved.

 Notice that because of  $T\dot{S}\doteq k_BT \dot{\cal{S}}$  we do have
\begin{equation}
 T\dot{S}_{int} = T\dot{S} - \dot{Q} \geq 0 \Rightarrow T\dot{S} \geq \dot{Q}\, .
\end{equation}
In view  of $
j = \rho v = {\frac{\rho }{m\beta }} [ f - k_BT \nabla \ln \rho ] \doteq  - {\frac{\rho }{m\beta }}\nabla \Psi
$  i.e.  $v= - (1/m\beta ) \nabla \Psi $ and  $f=-\nabla V$, we can  introduce
\begin{equation}\Psi = V + k_BT \ln \rho
\end{equation}
whose mean value stands for   the  Helmholtz free  energy of   the random  motion
\begin{equation}
F \doteq \left< \Psi \right> = U - T S \, .
\end{equation}
Here  $S \doteq k_B {\cal{S}}$ and an internal energy is $ U = \left< V\right>$.

Assuming that $\rho $  and
$\rho V v$  vanish at the integration volume boundaries we get
\begin{equation}
\dot{F}  =   \dot{Q} - T\dot{S} =  - (m\beta )
 \left<{v}^2\right> = - k_BT \dot{\cal{S}}_{int} \leq 0 \, . \label{helm}
\end{equation}
As long as there is a positive   entropy production, the Helmholtz free energy
decreases as a function of time  towards its minimum.
If there is none, the Helmholtz free energy remains constant.

With the  external forcing reintroduced, of particular interest is the  regime   $\dot{\cal{S}} =0$.
This occurs necessarily,
if the diffusion current  vanishes and   one   encounters
the state of equilibrium   with an  invariant density $\rho _{*}$.
 Then,    $b=u = D \nabla  \ln \rho _{*} $ and
$ -(1/k_BT)\nabla V = \nabla \ln\, \rho _{*}$ implies $\rho _{*} = {\frac{1}Z} \exp[ - V/k_BT]$.
Hence $\Psi _{*} = V + k_BT \ln \rho _{*}$ and therefore  $\langle \Psi _{*} \rangle =
 - k_BT \ln Z  \doteq  F_{*}$,  with   $Z= \int \exp(-V/k_BT) dx$,  is  a  minimum  of  the Helmholtz
free  energy $F$.

For the  free Brownian motion there is  no invariant density and we have $V=0 =b$, while $ v=-D\nabla \ln \rho = - u$, and therefore
   $ \dot{Q} =0  \Rightarrow  \dot{F} = - T\dot{S} = - m\beta D^2 \int [{\frac{(\nabla \, \rho )^2}{\rho }}]\, dx \leq 0 $.

\section{Entropy methods in  the  Schr\"{o}dinger picture  quantum dynamics}

A  pure state of the quantum system   and its  Schr\"{o}dinger picture dynamics
 are normally regarded    in conjunction with  the notion of a  thermodynamically \it isolated \rm quantum system.
A standard tool in the thermal context would be the von Neumann entropy notion which is known to vanish on pure states and
to be insensitive to the unitary quantum    evolution.
In below, we shall pay  attention to the Shannon entropy properties  in the quantum context\cite{gar}, to  demonstrate that a number of
 essentially thermodynamical features is encoded in the  apparently    non-thermodynamical
   regime of the Schr\"{o}dinger picture  quantum dynamics.

We come back to  the Schr\'{o}dinger  evolution of pure states in $L^2(R)$.
We employ  the natural boundary data (actually, the Dirichlet boundary conditions make the job) and
vanishing of various expressions at integration boundaries  is implicit, in all averaging procedures in below.
One must be aware that we pass-by a number of mathematical subtleties and take for granted that
various computational steps are allowed.

The continuity equation is a direct consequence of the Schr\"{o}dinger equation. It is less obvious
that, after employing the hydrodynamical velocity fields $u(x,t)$ and $v(x,t)$, the Fokker Planck equation for
 $\rho = |\psi |^2$ may be deduced. We have:
\begin{equation}
\partial _t\rho =
D\triangle \rho -  \nabla \cdot ( b \rho )
\end{equation}
where $b=v+u =\nabla (s+D\ln \rho )$ where $u=D\nabla \ln
\rho $.

The Shannon entropy of a continuous probability distribution ${\cal{S}} = -\langle \ln \rho \rangle $ follows and yields
\begin{equation}
  D \dot{\cal{S}}  =  \left< {v}^2\right>
    -  \left\langle {b}\cdot {v} \right\rangle \doteq D (\dot{\cal{S}}_{int}    +  \dot{\cal{S}}_{ext})
\end{equation}
which is a straightforward analog of the  $II^{nd}$ law of thermodynamics in the considered  quantum mechanical context:
\begin{equation}
\dot{\cal{S}}_{int} = \dot{\cal{S}} -  \dot{\cal{S}}_{ext}  =   (1/D)\left< {v}^2\right> \geq 0
 \Rightarrow \dot{\cal{S}}\geq \dot{\cal{S}}_{ext} \, .
\end{equation}
 To address an analog of the  $I^{st}$ law we need to translate
to the present setting the previously discussed thermodynamic
notions of
$U$ and $F = U-TS$, where the notion of \it temperature \rm is the most serious obstacle. We have no obvious notion  (nor physical intuitions  about)
   of the  temperature for
quantum systems in their pure states (for large molecules, like fullerenes or the likes, the notion of  an internal temperature makes
sense, but we aim to consider  any quantum system in a pure state, small or large).  Therefore, we shall  invoke a dimensional
artifice\cite{broglie}.

We formally introduce
\begin{equation}
  k_BT_0\doteq \hbar \omega _0\doteq mc^2
\end{equation}
   and thence
   \begin{equation}
D=\hbar/2m \equiv  k_BT_0/m\beta _0
\end{equation}
 with  $\beta _0 \equiv 2\omega _0 =
2mc^2/\hbar $, and so arrive at  the dimensionally acceptable identity
\begin{equation}
 k_BT_0\dot{\cal{S}}_{ext} = \dot{Q}\, .
 \end{equation}

In view of:
\begin{equation}
v=\nabla s =  b- u =  \nabla (s + D\ln \rho ) - D \nabla \ln  \rho \doteq
\end{equation}
$$
 - {\frac{1}{m\beta }}\nabla (V + k_BT_0\ln \rho )\doteq - {\frac{1}{m\beta _0}}\nabla \Psi \, ,
$$
where  the time-dependent potential
\begin{equation}
V = V(x,t) \doteq  - m\beta _0(s + D\ln \rho )
\end{equation}
 is defined to stay in a  notational  conformity with the standard   Smoluchowski process
 (Brownian motion in a conservative force field\cite{gar}) definition $b=-\nabla V/m\beta _0 $,
 we finally  get
\begin{equation}
- m\beta  \langle s\rangle \equiv \langle \Psi \rangle =\langle V \rangle - T_0 S  \Longrightarrow  F = U - T S  \, ,
\end{equation}
where $U= \langle V \rangle $ and $F=\langle \Psi \rangle $.

Remembering about an explicit time dependence of  $ b(x,t) =  -(1/m\beta _0)\nabla V(x,t)$, we  finally  arrive at
  the  direct analogue of the    $I^{st}$ law of thermodynamics in
the present quantum context:
\begin{equation}
\dot{U}=
\langle \partial _tV\rangle  - m\beta _0 \langle  b v \rangle =
\dot{W} + \dot{\cal{Q}}\, .
\end{equation}
The   term corresponding to the previous  "externally performed work"  entry reads $\dot{W}=\langle \partial _tV\rangle $.
But:
$$V= - m\beta  s  -k_BT\ln \rho  \,  \Longrightarrow  \,
\langle \partial _t V\rangle  = -  m\beta _0\langle \partial _t s\rangle  =\dot{W}$$
and therefore
\begin{equation}
- {\frac{d}{dt}} \langle s \rangle = -  \langle v^2\rangle - \langle \partial _t s \rangle  \Rightarrow
\dot{F} =  - T_0 \dot{S}_{int} + \dot{W}
\end{equation}
where $\dot{S}_{int}\geq 0$.

In view of Eq.~(\ref{total}), in the thermodynamical description
of the quantum motion, we encounter a  never vanishing constant
work term
\begin{equation}
  \dot{W} = m\beta _0{\cal{E}} = \beta _0\langle
 \hat{H}\rangle \, .
 \end{equation}
The quantum version  of the  Helmholtz-type  extremum principle reads:
\begin{equation}
\dot{F} - m\beta _0{\cal{E}}  = - T_0 \dot{S}_{int} \leq  0\, . \label{freeenergy}
\end{equation}
It is instructive to notice that
\begin{equation}
T_0\dot{S}_{int} = T_0\dot{S} - \dot{Q} \geq 0
 \Longleftrightarrow \dot{Q}\leq T_0\dot{S}
\end{equation}
 goes in parallel with
\begin{equation}
 \dot{F} \leq \dot{W} = \beta _0\langle \hat{H}\rangle \,  .
\end{equation}

Let us stress that the  non-vanishing  "external work" term is generic  to the quantum motion. If a stationary state is considered, our
$\langle \hat{H}\rangle $ is equal to  a corresponding energy eigenvalue.

 For negative eigenvalues,
the "work" term   corresponds   to what we   might possibly call the "work performed \it by \rm the system" (upon its, here  hypothetical,
 surrounding). Then $\dot{F}$ is negative and $F$  has  a chance to attain a minimum.

Since bounded from below Hamiltonians can  be replaced by positive operators, we may  in principle  view $m\beta _0{\cal{E}}=
\beta _0 \langle \hat{H}\rangle $ as a positive (constant and non-vanishing)  time rate of the
"work externally performed \it upon  \rm the system".   This observation encompasses   the case of positive energy spectra.
 Accordingly,  $\dot{F}$ may take both negative and  positive values. The latter   up to   an
 upper  bound $m\beta _0{\cal{E}}$.

 Basic temporal patterns of behavior, normally associated with  the non-equilibrium  thermodynamics of  \it closed \rm  irreversible   systems,
 somewhat surprisingly  have been   faithfully  reproduced in the  quantum Schr\"{o}dinger picture evolution which is  known to be  time-reversible.
 Nonetheless, we  have identified  direct analogues of  the  $I^{st}$  and the $II^{nd}$   laws  of thermodynamics, together with the involved
 notions of $\dot{S}_{int} \geq 0$ and $\dot{S}_{ext} = (1/T_0) \dot{Q}$. One should obviously remember about the pre-selected sense of time
 $t\in R^+$,  that was  employed in our discussion.

 An asymptotic $t\rightarrow \infty $ behavior of the quantum motion is controlled by  the  analog of the $II^{nd}$ law:
\begin{equation}
  \dot{F} - \dot{W} = -m\beta _0{\frac{d}{dt}}(\langle s\rangle +{\cal{E}} t )=
  - T_0 \dot{S}_{int} \leq 0 \, .  \label{work}
\end{equation}
  where there appears the    work (performed \it upon \rm or  performed \it by \rm  the system)
   term $\dot{W} = \langle \partial _t V\rangle = m\beta _0 {\cal{E}} $ value  whose sign is indefinite
   (either positive or negative).

Let us  recall  that in classical non-equilibrium thermodynamics  the  so-called  minimum entropy production
 principle\cite{glansdorf} is often invoked in connection with the "speed" with which
  a minimum  of the Helmholtz free energy is approached.
  For sufficiently large times, when the system is in the vicinity of
  the stationary (equilibrium) state, one expects that  the the entropy production $T \dot{S}_{int} \geq 0$  is
   a monotonically   decaying function of time, i.e. that
  \begin{equation}
{\frac{d}{dt}} \dot{S}_{int} < 0 \, .
\end{equation}

The quantum motion obviously looks different. In that case,  $\dot{F}$  may be positive and one cannot exclude
  transitions (including those of an oscillatory nature)  from negative to positive  $\dot{F}$ values and back.
In certain quantum states, the Helmholtz free energy $F$ may have a minimum, a maximum, an
 infinite number of local minima and maxima, or none at all. There is no reason for the  minimum entropy production principle
 to be valid in quantum theory, except for very special cases.

Since the work term is
  a constant of quantum motion,  we have :
 \begin{equation}
\dot{F}+ T_0\dot{S}_{int} = m\beta _0{\cal{E}}  \Longrightarrow   \ddot{F} = -  T_0 {\frac{d}{dt}} \dot{S}_{int} \, ,\label{feedback}
\end{equation}
which  formally   reproduces the temporal behavior characteristic to Smoluchowski diffusion processes, c.f. Eq.(\ref{helm}).
There  are  however  "speed" properties  which  are  special to the  quantum  dynamics   and have  no dissipative counterpart.

The above  time  rate formula  Eq.~(\ref{feedback}), which is  common  to both  quantum and diffusive motions scenarios, clearly is consistent with the
correlation of a minimum  of $\dot{F}$ with  a maximum of the  $\dot{S}_{int}$.
However,we  have as well  allowed  the reverse scenario  i.e. that a maximum of $\dot{F}$ may arise
  in conjunction with a minimum of $\dot{S}_{int}$.  More complicated, like e.g. oscillating,  forms of
   the  entropy production-Helmhholtz free energy
   interplay cannot be  a priori  excluded in the quantum case.

Remembering that $T_0\dot{S}_{int} = m\beta _0 \langle v^2\rangle$ and exploiting the total mean energy formula,
 Eqs. (\ref{total})  and(\ref{mean1}), we can identify the respective "speeds"  in conjunction with the Schr\"{o}dinger picture quantum motion.
  In  view of Eq.~(\ref{feedback}),
  the  pertinent time rates  stay in a  definite   \it negative  feedback  \rm  relationship.:
 \begin{equation}
  \ddot{F} =   + \beta _0  {\frac{d}{dt}} (  m \langle u^2\rangle +  2\langle {\cal{V}}\rangle )=
  -  m\beta _0 {\frac{d}{dt}} \langle v^2\rangle\, . \label{q}
\end{equation}

This observation  should be
contrasted with the behavior induced by
 diffusion-type processes, where
$Td \dot{S}_{int}/dt =   m\beta d
\langle v^2\rangle /dt $.  Now,  c.f. Eq.(\ref{mean}),  we have
\begin{equation}
 \ddot{F} = - \beta  {\frac{d}{dt}} (  m \langle u^2\rangle +  2\langle {\cal{V}}\rangle )=
-  m\beta  {\frac{d}{dt}} \langle v^2\rangle   \label{d}
\end{equation}
which really makes a difference (in view of the sign inversion in the functional expression for $\ddot{F}$). There is no feedback anymore.\\

\section{Outlook}

We have  discussed in detail the uncertainty/indeterminacy   measures  that can be   associated with  time-evolving
probability distributions of two basic origins.   We infer them  for  diffusion-type processes and  the
 Schr\"{o}dinger picture quantum dynamics.

 There are deep analogies between the quantum dynamics in the Madelung representation
 and the classical Fokker-Planck description of diffusion processes. We have exploited them in two   complementary ways.

First,  the  position-momentum indeterminacy relations   for diffusion processes  were deduced by  a modification of   major steps in
  the quantum procedure, compare e.g.  Eqs.~(\ref{ha}) and (\ref{ha1}).  Second, we have faithfully  reproduced in the quantum setting
  major thermodynamic relations between heat, work and free energy by adopting  to the quantum formalism  a number of  derivations that were
   consistently tested in the context of the Smoluchowski diffusion  processes.

The minor surprise is  that  major properties of   a   non-isolated but \it   closed \rm
 (we use the terminology of Ref.\cite{glansdorf})  random system
have been directly reproduced for the quantum system, which is  normally   considered as   thermally  \it isolated.   \rm
Our analysis allows to attribute to  the quantum system in a pure state major properties of a
non-isolated but \it closed \rm thermodynamical system.  The major difference between the quantum and diffusive behavior, if
restricted to thermodynamically motivated quantities, can be read out form the rate formulas Eqs.~(\ref{q}) and (\ref{d}).

To avoid misunderstandings, let us recall that in the classical situation work, heat and free energy have well defined meanings.
Work is interpreted as  an energy due to macroscopic degrees of freedom which perform an ordered motion and are perceived as
a source of work. Heat is  perceived as a thermal energy, while free energy is a maximal amount od energy which can be extracted as work.
A  physical meaning of these three concepts is definitely based on a  differentiation between the total system, the  investigated physical
 subsystem  and the  sources of work and heat (environment).

 We cannot propose  a clean physical picture  for deceivingly  thermodynamical patterns of behavior associated with the  quantum dynamics.
At the moment we have no satisfactory explanation of a possible physical meaning  of the "work performed upon" or "work performed by" the system,
nor heat, for an isolated quantum system. Albeit  we have demonstrated that this system shows up  patterns of behavior  that are
 characteristic for non-isolated \it closed \rm  thermodynamical systems, in parallel with those for diffusion-type processes.

On formal grounds, the present paper exploits properties of
$-\langle \ln \rho \rangle $ and of  $\langle (\nabla \ln \rho
)^2\rangle$, while admitting the time-dependence  of $\rho $.
 The functional $\langle (\nabla \ln \rho )^2 \rangle $ plays the major role in all our considerations and is responsible for the emergence of indeterminacy relations,
 both in the diffusive and quantum   motions.

 The final  outcomes of the discussion   do heavily  rely on the assumed factorization of
 the probability density $\rho $. It is accomplished either in terms of two real (time-conjugate)  functions $\rho = \theta \cdot \theta _*$,
  or in terms of two complex conjugate functions
 $\rho = \psi  \cdot \psi ^*$.      \\

\section{Appendix}

 Since the employed thermodynamic formalism may look strange for
 quantum theory practitioners, let us   exemplify the previous
 observations by invoking
 simple quantum motion cases.\\

  {\bf  Case 1:} {\it Free evolution}\\
 Let us consider  the probability density in one space dimension:
 \begin{equation}
 \rho (x,t) = {\frac{\alpha }{[\pi (\alpha ^4 + 4D^2t^2)]^{1/2}}}
 \exp \left( - {\frac{x^2\alpha ^2}{\alpha ^4 + 4D^2t^2}} \right)
 \end{equation}
 and the phase  function (we recall that $\psi = \rho ^{1/2} \exp(is/2D)$ is adopted)
 \begin{equation}
 s(x,t) = {\frac{2D^2x^2t}{\alpha ^4 + 4D^2t^2}} - D^2 \arctan
 \left( - {\frac{2Dt}{\alpha ^2}}\right)
 \end{equation}
 which determine  a free wave packet solution of the Schr\"{o}dinger  equation
 with the choice of  ${\cal{V}} \equiv 0$  and  the initial data $\psi (x,0)=
 (\pi \alpha ^2)^{-1/4} \exp(- x^2/2 \alpha ^2)$.

 One can readily deduce that
 \begin{equation}
  D (\dot{\cal{S}})_{int} =   \left< v^2\right> =
 \frac{8D^4t^2}{\alpha ^2(\alpha ^4  + 4D^2 t^2)}
 \end{equation}
 has an initial value $0$ and  attains a maximum $2D^2/\alpha ^2$ in  the  large time limit.
 Moreover, there holds
 \begin{equation}
 {\cal{E}}=  {\frac{1}2} (\langle v^2\rangle + \langle u^2 \rangle
 ) = {\frac{D^2}{\alpha ^2}} \,  ,
 \end{equation}
 where $\langle u^2 \rangle = (2D^2 \alpha ^2)/(\alpha ^4 +
 4D^2t^2)$. Clearly, an initial value of $\langle u^2\rangle $ is  $2D^2/\alpha ^2$,
 while $0$ stands for its asymptotic limit.

 The  feedback relationship Eq.~(\ref{feedback}) sets a link
 between the speed at which the entropy production attains its maximum and the speed at which $\dot{F}$ decreases
 towards its  minimal  value $\dot{F}^{min} = m\beta _0 {\cal{E}} - T_0\dot{S}_{int}^{max} $,
 compare e.g. Eq.~(\ref{freeenergy}).\\

 {\bf Case 2:} {\it  Steady state in a harmonic potential}\\
 We choose  a harmonic  potential ${\cal{V}}=  {\frac{1}2}\omega ^2 x$  in the Schr\"{o}dinger equation (\ref{Schroedinger})
  and  consider its solution with the probability density:
 \begin{equation}
 \rho (x,t) = \left( {\frac{\omega } {2\pi D}}\right) ^{1/2} \exp
 \left[ - {\frac{\omega }{2D}}\, \left( x - q(t)\right)^2 \right]
 \end{equation}
 and the phase function:
 \begin{equation}
   s(x,t) = (1/m)\left[ xp(t) -
 (1/2)p(t)q(t) - mD\omega t\right] \, ,
 \end{equation}
 where the classical harmonic dynamics with particle mass $m$ and
  frequency $\omega $ is involved such that
 $q(t)= q_0 \cos (\omega t) + (p_0/m\omega ) \sin (\omega t)$ and
 $p(t) = p_0\cos (\omega t) - m\omega q_0 \sin (\omega t)$.

 We have here  $v= \nabla s= p(t)/m$ and therefore:
 \begin{equation}
 D (\dot{\cal{S}})_{in} =  {\frac{p^2(t)}{m^2}}
 \end{equation}
 so that  in view of  $E/m = {\cal{E}} = p_0^2/2m^2 + \omega x_0^2/2
 + D\omega  $ and Eq.~(\ref{freeenergy}), remembering that
 $D=k_BT/m\beta_0= \hbar /2m$, we get
 \begin{equation}
 \dot{F} = \omega  k_BT_0 + m\beta _0 [ {\frac{p_0^2}{2m^2}}  +
 \omega {\frac{x_0^2}{2}} - {\frac{p^2(t)}{m^2}}]= \omega k_BT_0 +
 \beta _0[m\omega   {\frac{x^2(t)}{2}} -{\frac{p^2(t)}{2m}}] \, .
 \end{equation}
 It is interesting to observe that the actual behavior of $\dot{F}(t)$ depends on a difference of the potential
  and kinetic energies of the classical oscillator.\\

 {\bf Case 3:} {\it Stationary state}\\
 Let us make a brief comment on the case of stationary states. We  take a harmonic oscillator ground state as a reference.
  The entropy production vanishes, since $v=0$.  Then,  we have $\dot{F}=m\beta _0 {\cal{E}}_0= \beta _0 E_0$, where
  $E_0=\hbar \omega /2 =mD \omega $. Therefore
 \begin{equation}
   F(t)= (k_BT_0)\omega  t + const
 \end{equation}
    and  $F$ is a monotonically growing function.

  We recall that   presently  $\dot{F}=\dot{W}=\beta _0E_0$. The never ceasing time rate  of
  "work performed \it upon \rm  the surrounding"  needs to be kept in memory as a distinguishing feature of the quantum motion.

 Because of $-m\beta _0 \langle s \rangle = F$ and $\langle s \rangle = s$, we have
 \begin{equation}
 s(t)= - D \omega t + const \, ,
 \end{equation}
  as should be the case  in the exponent of the stationary  wave function $\psi = \rho ^{1/2}\exp(is/2D)$.
 Indeed,
  $-E_0t/2D = -\omega  t/2= s(t)/2D  - const$.\\


\begin{thebibliography}{99}
\bibitem{shannon} Shannon, C. E.: A mathematical theory of communication, {\it Bell Syst. Techn. J.}
 {\bf 27}, (1948), 379-423, 623-656
\bibitem{cover} T. M. Cover  and J. A.  Thomas,  {\it Elements of
Information Theory}, Wiley, NY, 1991
\bibitem{ohya} Ohya, M. and Petz, D., {\it Quantum Entropy and Its use},  Springer-Verlag, Berlin, 1993
\bibitem{stam} A. J.  Stam, Inf. and  Control, {\bf 2}, (1959), 101-112
\bibitem{dembo} A.  Dembo  and T. Cover,  IEEE Trans.
Inf. Th. {\bf 37}  (1991), 1501-1518
\bibitem{sobczyk} K. Sobczyk, Mechanical Systems and Signal Processing, {\bf 15} (2001), 475-498
\bibitem{gar} P. Garbaczewski, J. Stat. Phys. {\bf 123}, (2006),
315-355
\bibitem{gar1} P. Garbaczewski, Appl. Math \& Information Science,
{\bf 1}, (2007), 1-12
\bibitem{mackey} M. C. Mackey, M. Tyran-Kami\'{n}ska, Physica  {\bf A 365}, 360-382, (2006)
\bibitem{furth} R. F\"{u}rth, Zeitschr. Phys. {\bf 81}, (1933), 143-162
\bibitem{viola} F. Illuminati and L. Viola, J. Phys. A: Math. Gen. {\bf 28}, (1995), 2953-2961
\bibitem{golin} S. Golin, J. Math. Phys. {\bf 26}, 1985, 2781-2783
\bibitem{mycielski} I. Bia{\l}ynicki-Birula and J.  Mycielski,  Commun. Math. Phys.  {\bf 44}, (1975), 129-132
\bibitem{gar2} P. Garbaczewski, Rep. Math. Phys.   {\bf 56}, (2005), 153-160
\bibitem{hall}  M. J. W. Hall, Phys. Rev. {\bf A 62}, 012107, (2000)
\bibitem{shin} E. I. Verriest and D-R. Shin, Int. J. Theor. Phys. {\bf 32}, (1993), 333-345
\bibitem{zambrini} J-C. Zambrini, J. Math. Phys. {\bf 27}, (1986), 2307-2330
\bibitem{glansdorf} P. Glansdorf and I. Prigogine, {\it Thermodynamic Theory of Structure, Stability
and Fluctuations},  (Wiley, NY, 1971)
\bibitem{kondepudi}  D. Kondepudi and I. Prigogine, {\it Modern Thermodynamics}, Wiley, NY, 1998
\bibitem{hasegawa} H. Hasegawa, Progr. Theor. Phys. { 57}, 1523-1537, (1977)
\bibitem{qian} H. Qian, Phys. Rev. {\bf E 65}, 021111, (2002)
 \bibitem{shizume} K. Shizume, Phys. Rev. {\bf E 52}, 3495-3499, (1995)
\bibitem{broglie} L. de Broglie, {\it La Thermodynamique de la particule isol\'{e}e}, Gauthier-Villars, Paris, 1964
\end{thebibliography}
\end{document}